\documentclass{PoS}

\usepackage{xspace}
\usepackage{tabularx}
\newcolumntype{M}[1]{>{\centering\arraybackslash\hspace{0pt}}m{#1}}
\newcommand{\tev}{\ensuremath{~\mathrm{TeV}\xspace}}
\newcommand{\gev}{\ensuremath{~\mathrm{GeV}\xspace}}
\newcommand{\plots}[4]{
  \centering\newdimen\lwidth\newdimen\rwidth
  \settowidth{\lwidth}{\includegraphics[#1]{plots/#2}}
  \settowidth{\rwidth}{\includegraphics[#3]{plots/#4}}
  \begin{tabular}{M{\lwidth}M{\rwidth}}
    \includegraphics[#1]{plots/#2} & \includegraphics[#3]{plots/#4} \\
  \end{tabular}
}

\title{WG1: Structure Functions and Parton Densities}

\ShortTitle{WG1: Structure Functions and Parton Densities}

\author{\speaker{Lucian Harland-Lang}\\
  Rudolf Peierls Centre for Theoretical Physics, Oxford, UK\\
  E-mail: \email{lucian.harland-lang@physics.ox.ac.uk}}
\author{Philip Ilten\\
  School of Physics and Astronomy, University of Birmingham, Birmingham, UK\\
  E-mail: \email{philten@cern.ch}}
\author{Jan Kretzschmar\\
  Oliver Lodge Laboratory, University of Liverpool, Liverpool, UK\\
  E-mail: \email{jan.kretzschmar@cern.ch}}

\abstract{This paper gives a summary of selected highlights discussed in the
working group on ``Structure Functions and Parton Densities'' (WG1) at
the DIS 2017 conference. From the many talks presented we extract some
general themes discussed with respect to global PDF fits, new
PDF-sensitive measurements from the LHC experiments and elsewhere,
exploitation of new ideas, tools to perform PDF fits, PDFs of heavy
nuclei, and finally progress in basic theory calculations.
}

\FullConference{XXV International Workshop on Deep-Inelastic Scattering and Related Subjects\\ 3-7 April 2017\\ University of Birmingham, UK}

\begin{document}

\section{Introduction}

The ``Structure Functions and Parton Densities'' working group (WG1)
of the DIS 2017 conference featured 45 talks spanning a wide range of
interesting topics. The talks were organised in 10 sessions, including
one joint session with ``Physics with Heavy Flavours''
(WG5)~\cite{Giammanco:2017bvx} and two joint sessions with ``Hadronic
and Electroweak Observables'' (WG4)~\cite{wg4summary}. In the
following we summarise some selected highlights of these sessions.

\section{Global PDF Fits}

Updates from three major global fitting collaborations CT~\cite{CT},
MMHT~\cite{MMHT}, and NNPDF~\cite{NNPDF} were presented. In all cases,
the results shown made use of an increasingly wide range of data from
the LHC experiments and the LHC data are playing an increasingly
important role in global PDF fits. Studies beyond the new LHC data
were also presented, for example regarding updates on the photon PDF
from the CT and MMHT groups and intrinsic charm by the CT and NNPDF
groups, as discussed further below.  Work towards new public releases
of PDF sets is ongoing for the CT and MMHT groups, while the updated
NNPDF3.1 set based on the work presented here is now
available~\cite{Ball:2017nwa}.

\begin{figure}[tb]
  \plots{height=0.32\columnwidth}{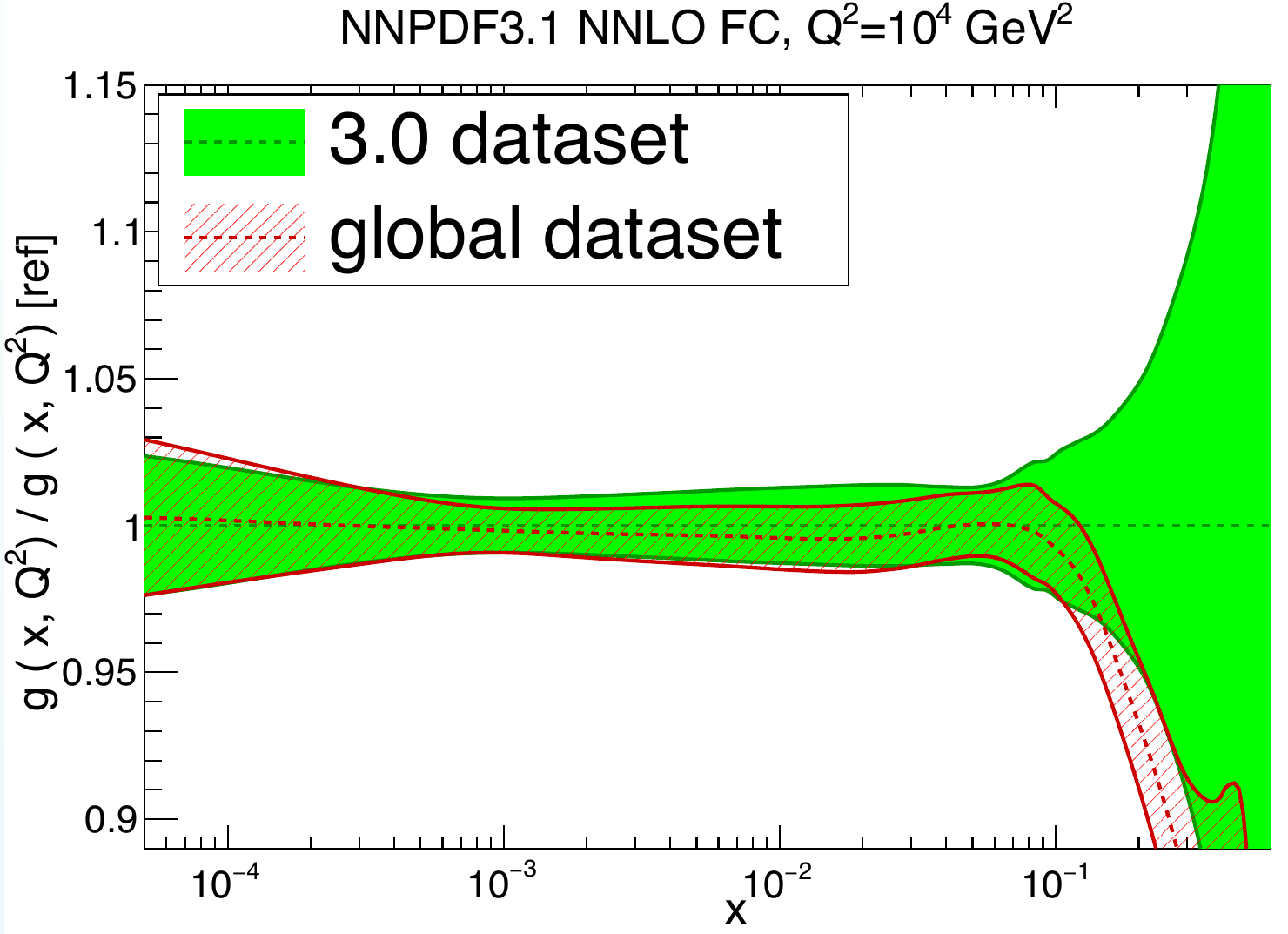}
        {height=0.32\columnwidth}{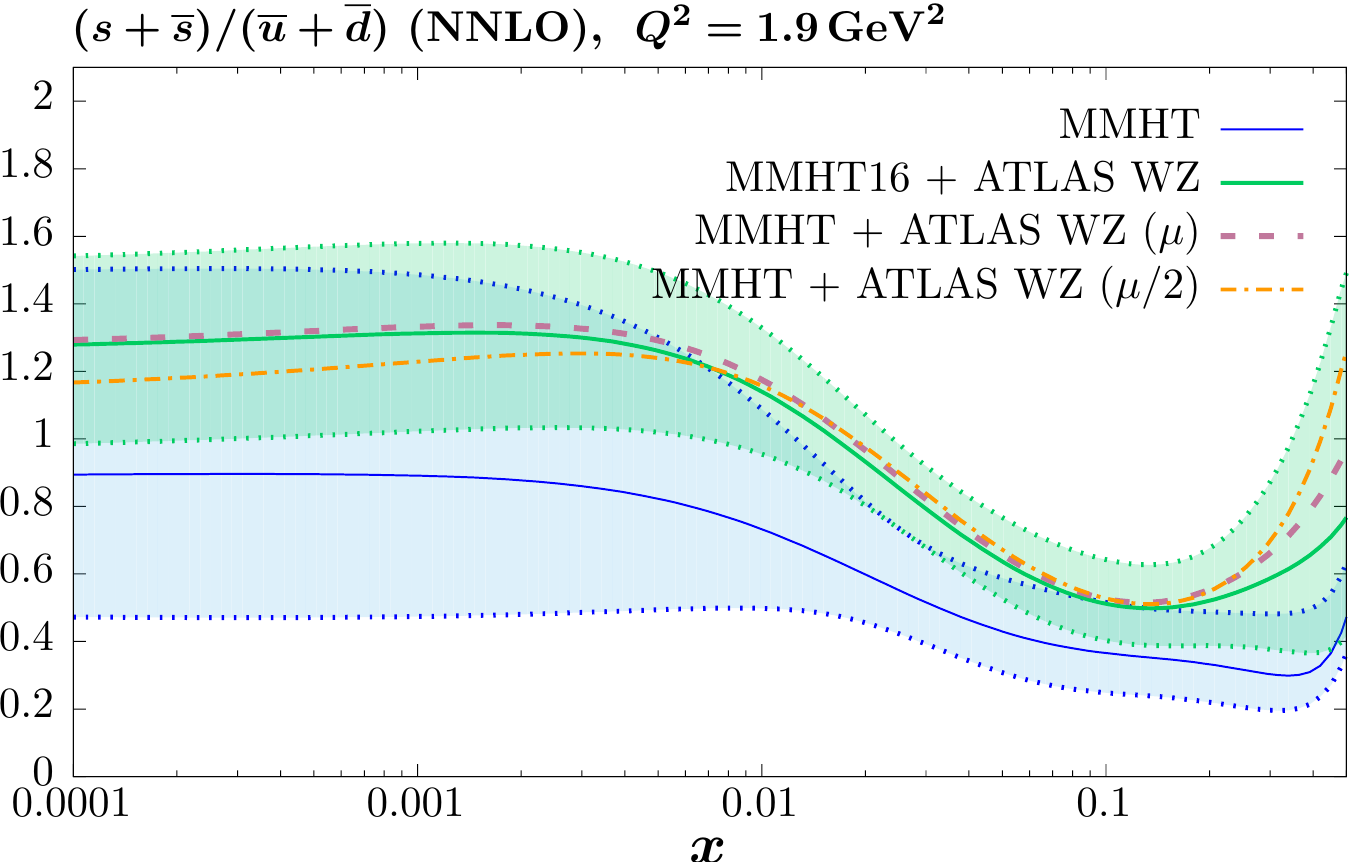}
  \caption{(Left) Impact of new data on the gluon PDF, predominately
    from the LHC, in the new NNPDF3.1 fit as compared to the previous
    NNPDF3.0 analysis~\cite{NNPDF}. (Right) Impact of the high
    precision $7\tev$ ATLAS $W$ and $Z$ production data on
    the strange-quark contribution to the proton in the MMHT
    fit~\cite{MMHT}.\label{fig:pdfs}}
\end{figure}

In Fig.~\ref{fig:pdfs} (left) the impact of new data on the gluon,
predominately from the LHC, is shown by comparing the new
determination to the previous NNPDF3.0 determination; the changes are
significant, especially at larger $x$. A similar effect is observed by
CT and MMHT. In the latter case, for example, LHC data is found to
provide the most important constraints in the latest fit for 21 out of
the 50 eigenvector directions. In Fig.~\ref{fig:pdfs} (right) the
constraint that the new high precision ATLAS $W$ and $Z$ production
data~\cite{Sommer,Aaboud:2016btc} places on the strange-quark
distribution in the proton (the ratio of the strange to the
light-quark sea is shown here) is demonstrated, within the MMHT
fit. The impact is large, with the strange-quark fraction found to be
somewhat higher than the previous determination, although consistent,
and the uncertainties are reduced. A similar trend is seen in the
NNPDF analysis.

LHC data certainly represent important opportunities to improve the
knowledge on PDFs. However, there are many challenges for the PDF
fitters to include such increasingly precise data effectively within a
PDF fit. Some of these challenges were also highlighted in this
conference, and will be discussed below.

\section{New Measurements sensitive to PDFs and $\alpha_S$}

There has been significant progress in the past year regarding new
experimental measurements. The majority of new results were presented
by the ATLAS, CMS, and LHCb collaborations at the LHC, but new results
are still being made public from the HERA experiments like H1. These
data can be used to further constrain PDFs and measure the strong
coupling $\alpha_S$. The results span a wide range of physical
observables including jets, vector bosons in association with heavy
flavour jets, and high-precision inclusive vector-boson measurements.

\begin{figure}[tb]
  \plots{height=0.51\columnwidth}{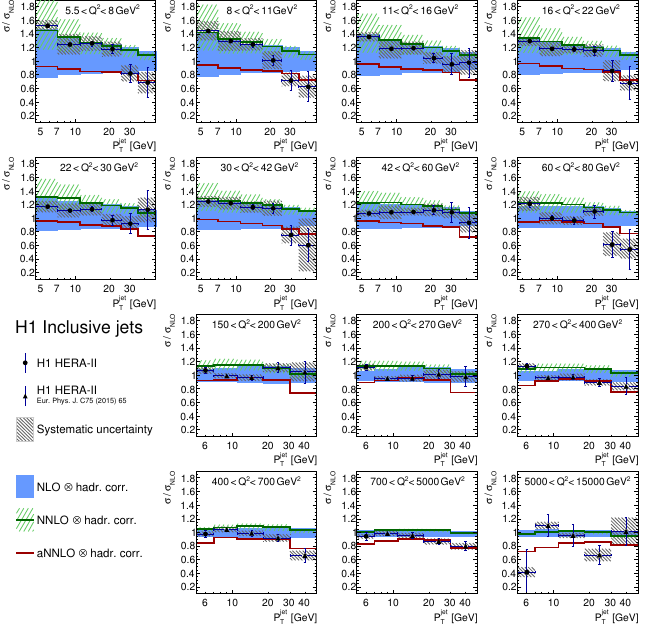}
        {height=0.45\columnwidth}{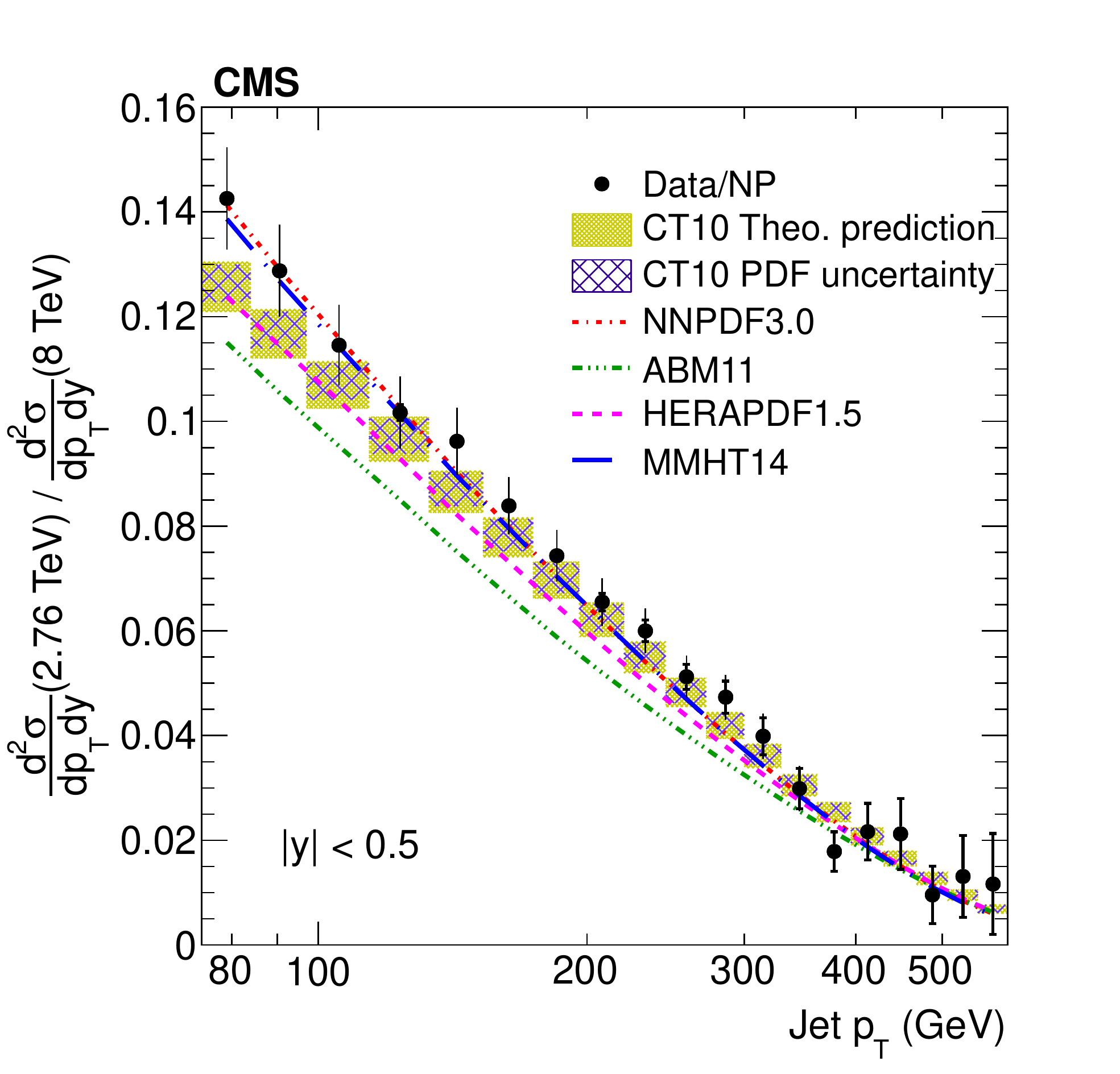}
   \caption{(Left) Inclusive differential jet cross sections from
     H1~\cite{Andreev:2016tgi}. (Right) Ratio between the inclusive
     jet cross sections at $\sqrt{s} = 2.76$ and $8\tev$ from
     CMS~\cite{Khachatryan:2016mlc}.\label{fig:jets}}
\end{figure}

Double-differential jet cross sections in DIS have been measured as a
function of virtuality $Q^2$ and jet $p_\mathrm{T}$ by the H1
collaboration~ \cite{Andreev:2016tgi, Britzger}.  These data have been
compared to the latest next-to-next-to-leading order (NNLO) theory
calculations for the first time for a $Q^2$ range of $5.5$ to
$15000\gev^2$ and a jet $p_\mathrm{T}$ range of $5$ to $40\gev$. As
seen in Fig.~\ref{fig:jets} (left), the NNLO predictions provide a
better description of the cross section shape as compared to NLO
predictions, as well as a reduction in the theoretical uncertainties
as estimated by scale variations. Double-differential inclusive jet
cross sections in $pp$ collisions at $8\tev$ in jet $y$ and
$p_\mathrm{T}$ have been performed by the ATLAS and CMS
collaborations~\cite{Dandoy,Lipka,Eren,Aaboud:2017dvo,Khachatryan:2016mlc}. The
CMS result also provides ratios between the cross sections measured at
$\sqrt{s} = 8$ and $2.76\tev$~\cite{Khachatryan:2016mlc}, resulting in
a partial cancellation of systematic uncertainties and greater
sensitivity to PDFs. As shown in Fig.~\ref{fig:jets} (right), the
HERAPDF1.5 and MMHT14 PDF sets describe this ratio well, while the
ABM11, CT10, and NNPDF3.0 PDF sets systematically under-estimate the
cross section ratio at lower jet $p_\mathrm{T}$. Analyses of the ATLAS
measurement at $\sqrt{s} = 8\tev$~\cite{Aaboud:2017dvo} within the
framework of global PDF fits have reported difficulties in describing
the measurement across all jet $y$ bins, similar to the previous
$\sqrt{s} = 7\tev$ ATLAS measurement~\cite{Aad:2014vwa}. However,
decorrelating two of the jet energy scale systematics has been found
to significantly improve the description of the $7\tev$
data~\cite{MMHT,Harland-Lang:2017ytb}.  Further studies of the $8\tev$
data are therefore required to fully assess its impact on PDFs.

Differential cross section measurements for dijet and trijet events
have also been made by the H1 and CMS
collaborations~\cite{Andreev:2016tgi,
  Britzger,Lipka,Eren,Sirunyan:2017skj,Khachatryan:2016mlc}. The H1
dijet results are again compared to NNLO theory predictions. While the
NNLO predictions describe the shape well, they consistently
overestimate the cross section. The H1 trijet measurement is more
limited by statistical uncertainties and found to be in good agreement
with the NLO predictions. A ratio between the CMS inclusive trijet and
dijet cross sections at $\sqrt{s} = 8\tev$ as a function of the
average $p_\mathrm{T}$ of the two highest $p_\mathrm{T}$ jets in the
event is particularly sensitive to
$\alpha_s$~\cite{Lipka,CMS-PAS-SMP-16-008}. Fits of this ratio, as
well as the dijet and trijet cross sections, were performed using a
variety of PDF sets to determine $\alpha_S(M_z)$. The extraction of
$\alpha_S(M_Z)$ from the ratio fit typically resulted in lower values
of $\alpha_S(M_z)$, giving $0.1139 \pm 0.0032$ with CT14, than the
simultaneous dijet and trijet cross section fits, which gives $0.1161
\pm 0.0029$ with CT14, where the quoted uncertainties exclude QCD
scale variations.  Electroweak corrections were calculated for the
dijet fits but not the trijet fits. However, these corrections are
assumed to have a minor effect in the cross section ratio, and so the
$\alpha_s(M_Z)$ value extracted from the ratio is considered as the
nominal result in this analysis, leading to the central result of
$\alpha_S(M_z) = 0.1150 \pm 0.0023 ^{+0.0050}_{-0.0000}$ using the
MSTW2008 PDF set without LHC jet data. The second uncertainty is from
QCD scale variations while the the first includes all other remaining
uncertainties.

\begin{figure}[tb]
  \plots{height=0.38\columnwidth}{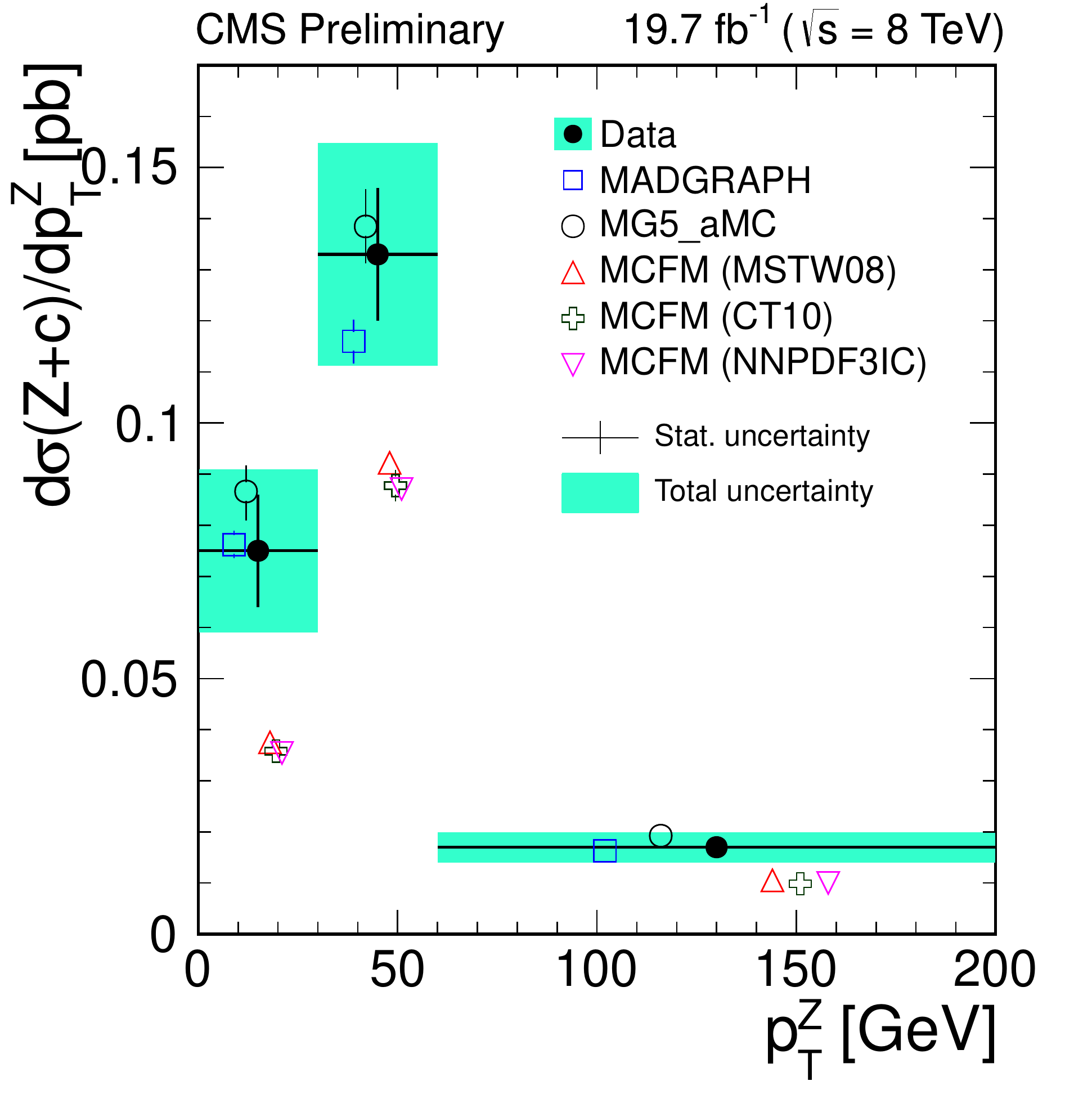}
        {height=0.38\columnwidth}{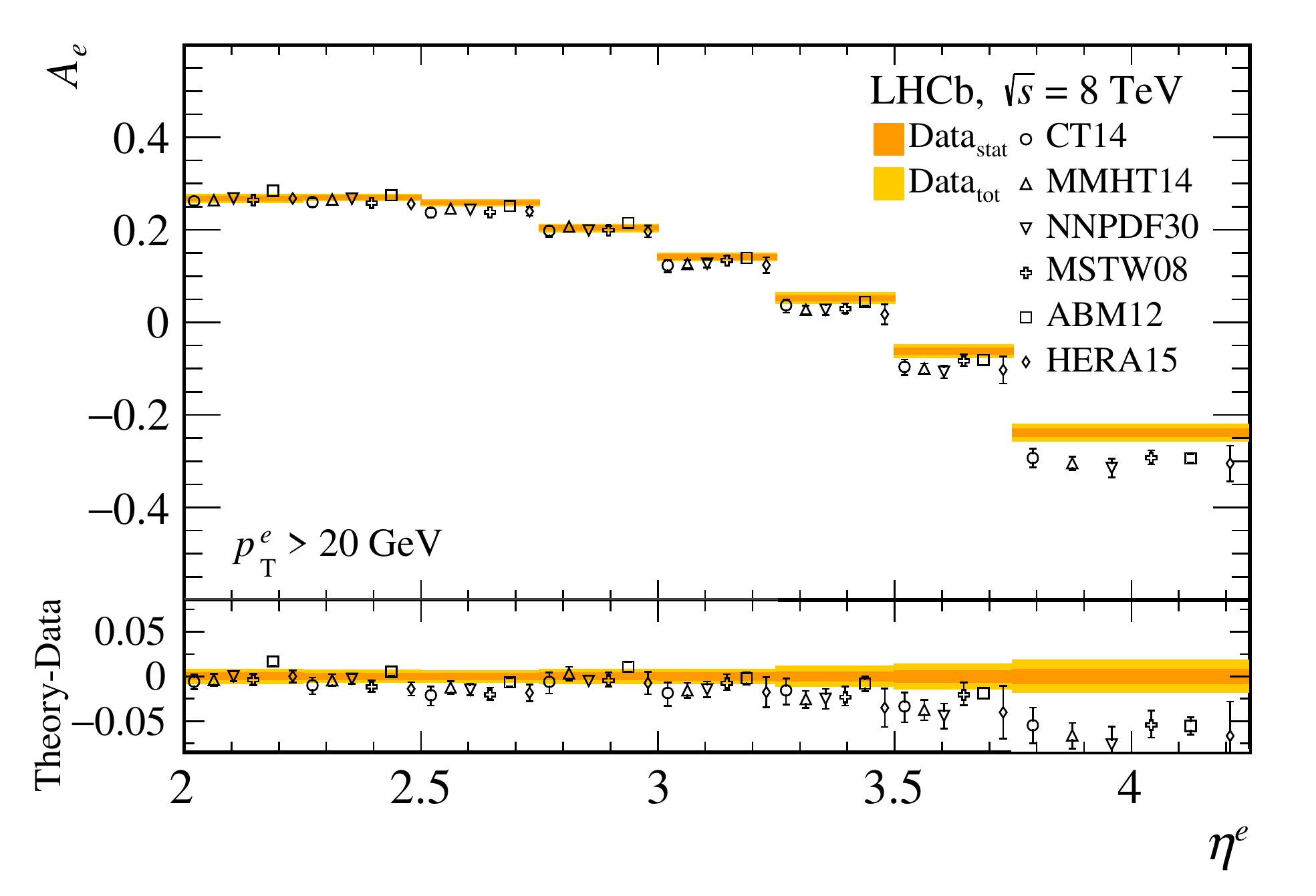}
  \caption{(Left) Differential $Z + c$ cross section measured by
    CMS~\cite{Sirunyan:2017pob}. (Right) Charge asymmetry for $W \to
    e$ measured as a function of electron pseudorapidity from
    LHCb~\cite{Aaij:2016qqz}.\label{fig:ewk1}}
\end{figure}

Larger datasets and progress in tagging of heavy-flavour jets has
allowed new measurements of vector bosons in association with heavy
flavour jets from both CMS and LHCb to be made. The CMS collaboration
presented differential $Z+c$ cross sections measured at $\sqrt{s} =
8\tev$ as a function of the $p_\mathrm{T}$ for both the $Z$-boson and
$c$-jet~\cite{Roland, Sirunyan:2017pob}.  This is in principle
sensitive to intrinsic charm contributions to the PDFs, but so far
large statistical and $c$-tagging efficiency uncertainties do not
allow for strong constraints. The differential $Z+c$ and ratio of
$Z+c$ to $Z+b$ cross sections are compared to theory predictions from
\textsc{MadGraph}, aMC@NLO, and MCFM in Fig.~\ref{fig:ewk1}. While the
\textsc{MadGraph} and aMC@NLO predictions are found to be in good
agreement with data, the MCFM predictions are consistently lower than
the measured result, independent of the PDF set used, and further work
is required to fully understand the implications of this
measurement. A combined measurement of the $W + c\bar{c}$, $W +
b\bar{b}$, and $t \bar{t}$ cross sections was performed by
LHCb~\cite{Aaij:2016vsy} using $\sqrt{s} = 8\tev$ data with an
electron or muon final state in association with two jets tagged to
contain heavy-flavour hadrons. While this measurement is statistically
limited, future updates will provide valuable information on the
strange-quark PDF.  Further measurements of the $W+c$ final state by
the ATLAS, CMS, and LHCb collaborations will help constrain the size
and possible asymmetry of the strange-quark PDF, and resolve the
current mild tensions between different datasets.

New precise differential measurements of inclusive vector-boson
production have been made by the ATLAS, LHCb, and STAR
collaborations. Integrated and differential $W$ and $\gamma^*/Z$ cross
sections at $\sqrt{s} = 7\tev$ were measured by the ATLAS
collaboration~\cite{Sommer, Aaboud:2016btc} using both electron and
muon final states with systematic uncertainties well below the percent
level and negligible statistical uncertainty, testing lepton
universality at the percent level. A PDF analysis was performed using
these data combined with HERA data, and the strange-to-light sea-quark
ratio was found to be near unity, a result confirmed in global PDF
analyses performed by MMHT~\cite{MMHT} and NNPDF~\cite{NNPDF}, albeit
with a reduced fit quality.  The MMHT study and an ATLAS study
demonstrated that halving the nominal factorisation and
renormalisation scale, the di-lepton mass, improves the fit quality
but has only a small impact on the strange-quark
PDF~\cite{MMHT,Aaboud:2016btc}. New ATLAS measurements of high-mass
Drell-Yan cross sections were presented in~\cite{Zinser,Aad:2016zzw},
which are generally well described by different PDF sets, but were
found to reduce significantly the variations of the photon PDF in the
NNPDF2.3QED set.

\begin{figure}[tb]
  \plots{width=0.5\columnwidth}{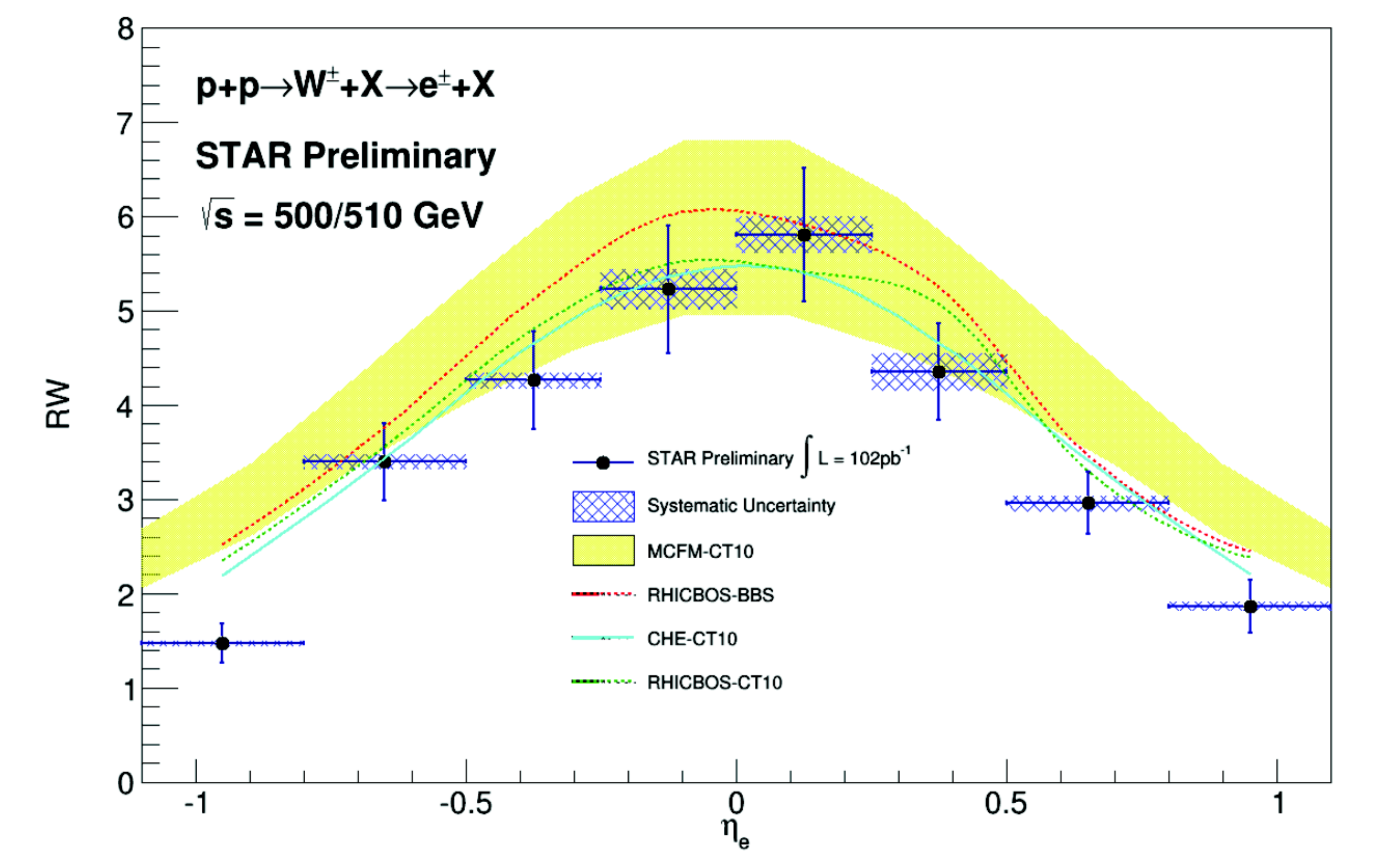}
        {width=0.48\columnwidth}{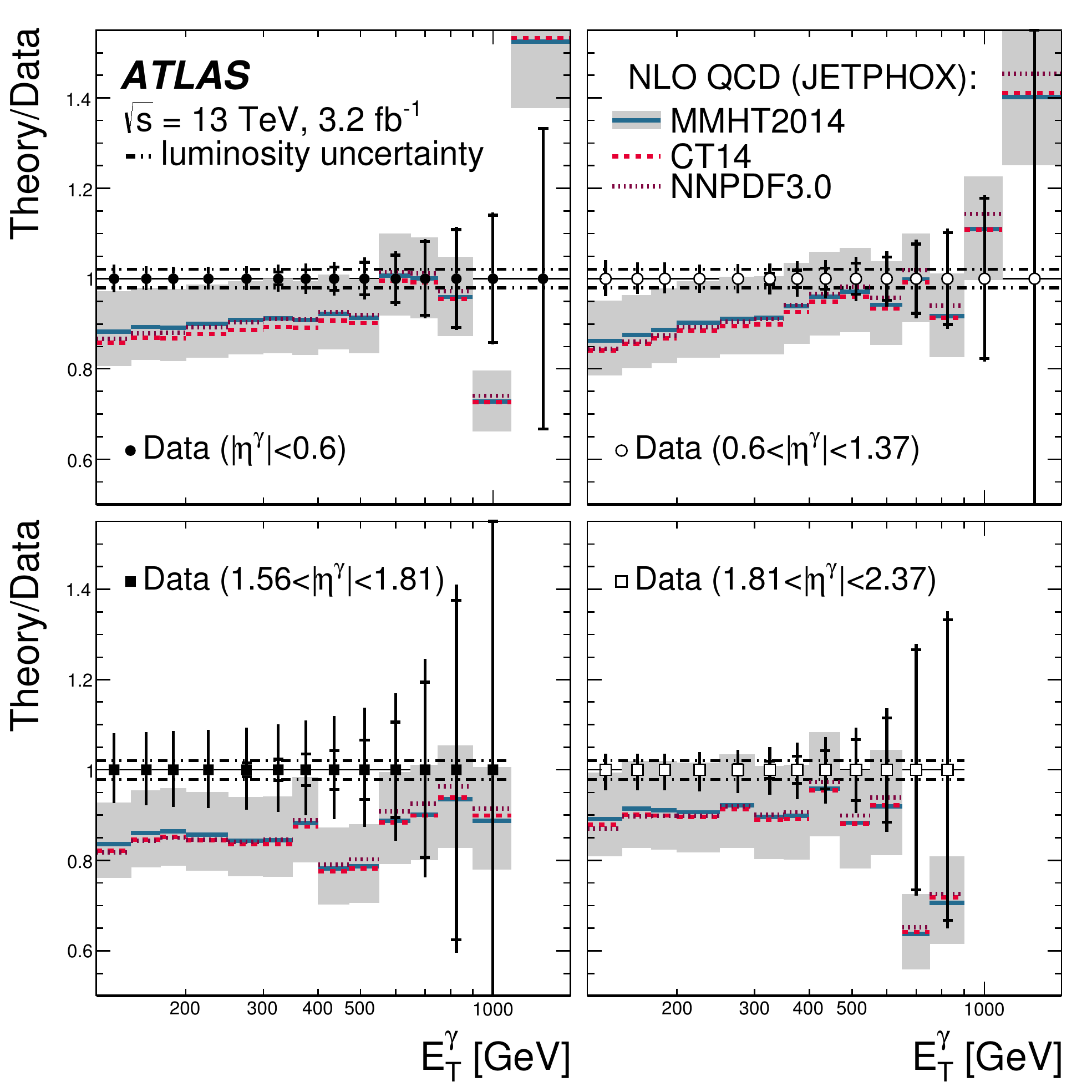}
  \caption{(Left) Differential $W^+/W^-$ cross section ratio from STAR
    as a function of electron pseudorapidity. (Right) Isolated photon
    cross section from ATLAS compared to NLO theory
    predictions.\label{fig:ewk2}}
\end{figure}

Similar precision measurements of $\gamma^*/Z$ at $\sqrt{s} = 13\tev$
and $W$ at $\sqrt{s} = 8\tev$ were performed by
LHCb~\cite{Kucharczyk,Aaij:2016qqz,Aaij:2016mgv}, which is sensitive
to PDFs at both lower and higher $x$ due to the forward rapidity
coverage of the detector.  These results were found to be in good
agreement with NNLO predictions using a variety of PDF sets. However,
the charge asymmetry at forward lepton pseudorapidity was found to be
systematically underestimated by all PDF sets tested, see
Fig.~\ref{fig:ewk1} (right). The $W^+/W^-$ ratio using $W$-boson
decays into an electron final state was measured differentially in
pseudorapidity by the STAR collaboration using RHIC $pp$ data at
$\sqrt{s}$ between $500$ and $510\gev$~\cite{Fazio}.  The results are
shown in Fig.~\ref{fig:ewk2} (left), where a slight tension between
theory and data at large and small electron pseudorapidity is visible.

Finally, measurements of isolated photon production differential in
photon $E_\mathrm{T}$ and pseudorapidity, were performed by ATLAS at
$\sqrt{s} = 13\tev$~\cite{Glasman,Aaboud:2017cbm}. The NLO predictions
were found to systematically underestimate the photon cross section,
particularly at lower $E_\mathrm{T}$ as shown in Fig.~\ref{fig:ewk2}
(right), demonstrating the need for NNLO predictions that have
recently become available~\cite{Campbell:2016lzl}.

\section{New Ideas to constrain PDFs}

While many of the discussions in WG1 were centred around existing fit
frameworks and the inclusion of data traditionally used to constrain
PDFs, many new ideas and approaches were also discussed. Novel
possibilities to constrain the gluon PDF at low $x$ were introduced
in~\cite{Rojo} and~\cite{Sjones}. In~\cite{Rojo,Gauld:2016kpd}, data
of open-charm hadron production from $pp$ collisions in the forward
region taken by the LHCb experiment was demonstrated to have
significant constraining potential. These constraints have
implications for the production of neutrinos though heavy-flavour
production in ultra high-energy cosmic ray collisions and for physics
at a possible $\sqrt{s} = 100\tev$ FCC. In~\cite{Sjones,Jones:2016icr}
the exclusive photoproduction of heavy quarkonia ($J/\psi$,
$\Upsilon$), where both protons remain intact after the collision, was
also shown to have a significant impact on the gluon PDF in the very
low $x$ region. However, both processes are affected by large
theoretical uncertainties estimated from varying the factorisation and
renormalisation scales. In~\cite{Rojo} the impact was minimised by
choosing suitably normalised observables and assuming that these
variations are largely correlated, while in~\cite{Sjones} a physically
motivated scale choice was taken.

\begin{figure}[tb]
  \plots{height=0.35\columnwidth}{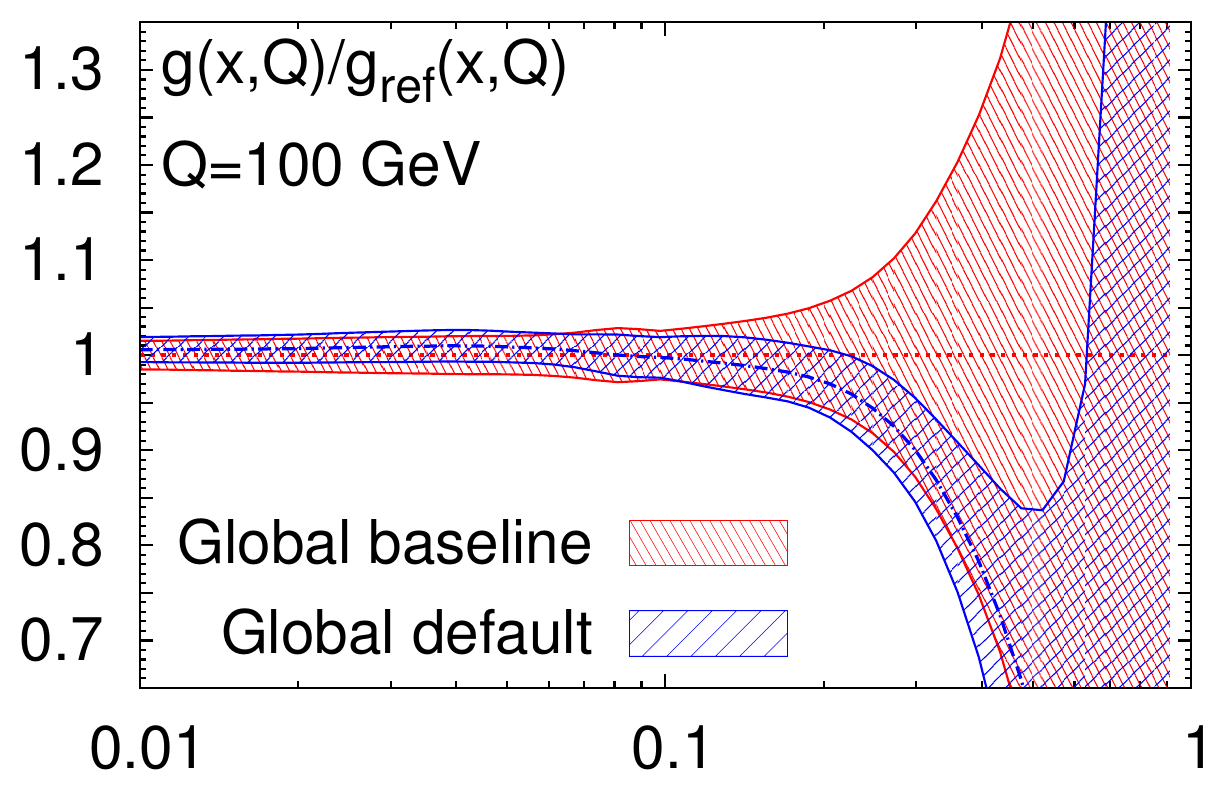}
        {height=0.35\columnwidth}{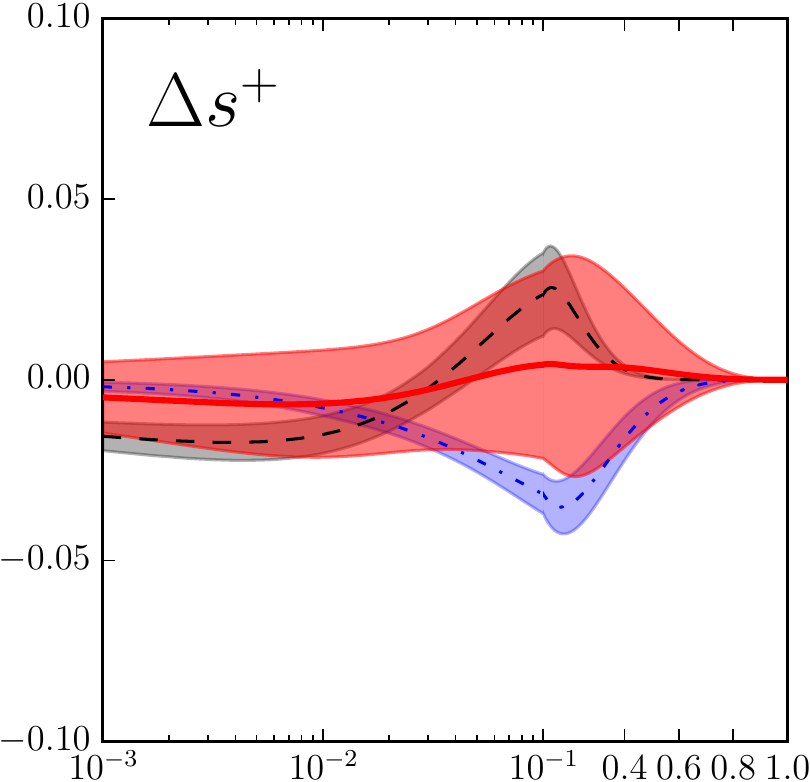}
  \caption{(Left) Impact on the gluon PDF with respect to the global
    baseline of ATLAS and CMS differential top production data at
    NNLO~\cite{Czakon:2016olj}. The ``optimal'' fit corresponds to the
    recommended choice of observables described in the text. (Right)
    The polarised strangeness $\Delta s^+$ distribution extracted from
    the JAM17 preliminary combined fit to polarised PDFs and
    fragmentation functions~\cite{Ethier}.\label{fig:ttbarjam}}
\end{figure}

The high $x$ gluon was considered in two separate contributions,
\cite{Noceratop} and \cite{Ubiali}. In~\cite{Noceratop} a PDF fit was
described that utilises the new NNLO QCD theory calculation combined
with the recent ATLAS and CMS data for differential top-quark
production; these measurements are differential in
$m_{t\overline{t}}$, $y_t$, $y_{t\overline{t}}$ and $p_\mathrm{T}^t$, see
~\cite{Czakon:2016olj} for details. Although there are tensions in
some distributions between the ATLAS and CMS data, the impact on the
gluon distribution of the individual distribution choices was mostly
consistent. A recommendation was made to use a combination of the
normalised top and top pair rapidity distributions, for which no
tension was apparent. Additionally, the sensitivity in the high
$m_{t\overline{t}}$ region, where beyond-the-standard-model effects
may enter, is smaller. A clear impact on the gluon PDF at high $x$,
comparable to the effect of inclusive jet data, was demonstrated, as
shown in Fig.~\ref{fig:ttbarjam} (left). The CMS collaboration has
also recently explored the impact of double-differential $t\bar{t}$
measurements as presented in~\cite{Lipka,Sirunyan:2017azo}, although
here the theoretical calculation is only available at NLO so
far. ATLAS presented a study that correlates $t\bar{t}$ and $Z$
production in $pp$ collisions at various $\sqrt{s}=7,8,13~\tev$ to
improve the impact on PDF constraints~\cite{Zinser,Aaboud:2016zpd}.

In~\cite{Ubiali,Nocera:2017zge} the impact of measurements of the
$Z$-boson transverse momentum distribution on the determination of
PDFs was explored. The theoretical calculations are now available at up
to NNLO in QCD and NLO in EW~\cite{Boughezal:2017nla}. To achieve a
good description of the data, an additional $1\%$ of bin-to-bin
uncorrelated uncertainty was included, which it was argued was
necessary to account for statistical uncertainties on the theory
calculations and a possible underestimation of experimental
uncertainties. The impact of these ad-hoc uncertainties on the
extracted PDFs was found to be relatively minor. The analysis found
some tension when including the normalised distributions,
\textit{e.g.} the ATLAS $7\tev$ data. This is believed to be a feature
of either the treatment of the experimental correlations or the
theory, in the case of normalised distributions, where both the high
$p_\mathrm{T}^Z$ region and the integral over the full distributions enter
the fit. A fit to the $8\tev$ unnormalised distributions was found to
give a significant reduction in PDF uncertainties in comparison to a
HERA-only PDF set.

Progress in the determination of polarised PDFs was also
presented. In~\cite{Ethier,Ethier:2017zbq} the first simultaneous
extraction of both polarised PDFs and fragmentation functions from
DIS, semi-inclusive DIS (for which the pion and kaon fragmentation
functions are required), and semi-inclusive annihilation data at NLO
within the JAM17 framework was discussed. To achieve this, an
iterative Monte Carlo fitting method was used. A good description of
the data was achieved, and as shown in Fig.~\ref{fig:ttbarjam}
(right), a $\Delta s^+$ consistent with zero across the entire $x$
range within fairly large uncertainties was found, in contrast to some
earlier studies. In~\cite{Maji,Maji:2017bcz} a theoretical calculation
of the transverse momentum dependent PDFs within the light-front quark
di-quark model was presented. The resultant PDFs were found to be
consistent with global fits, and were able to successfully describe
the Collins asymmetry in semi-inclusive DIS.

\begin{figure}[tb]
  \plots{height=0.35\columnwidth}{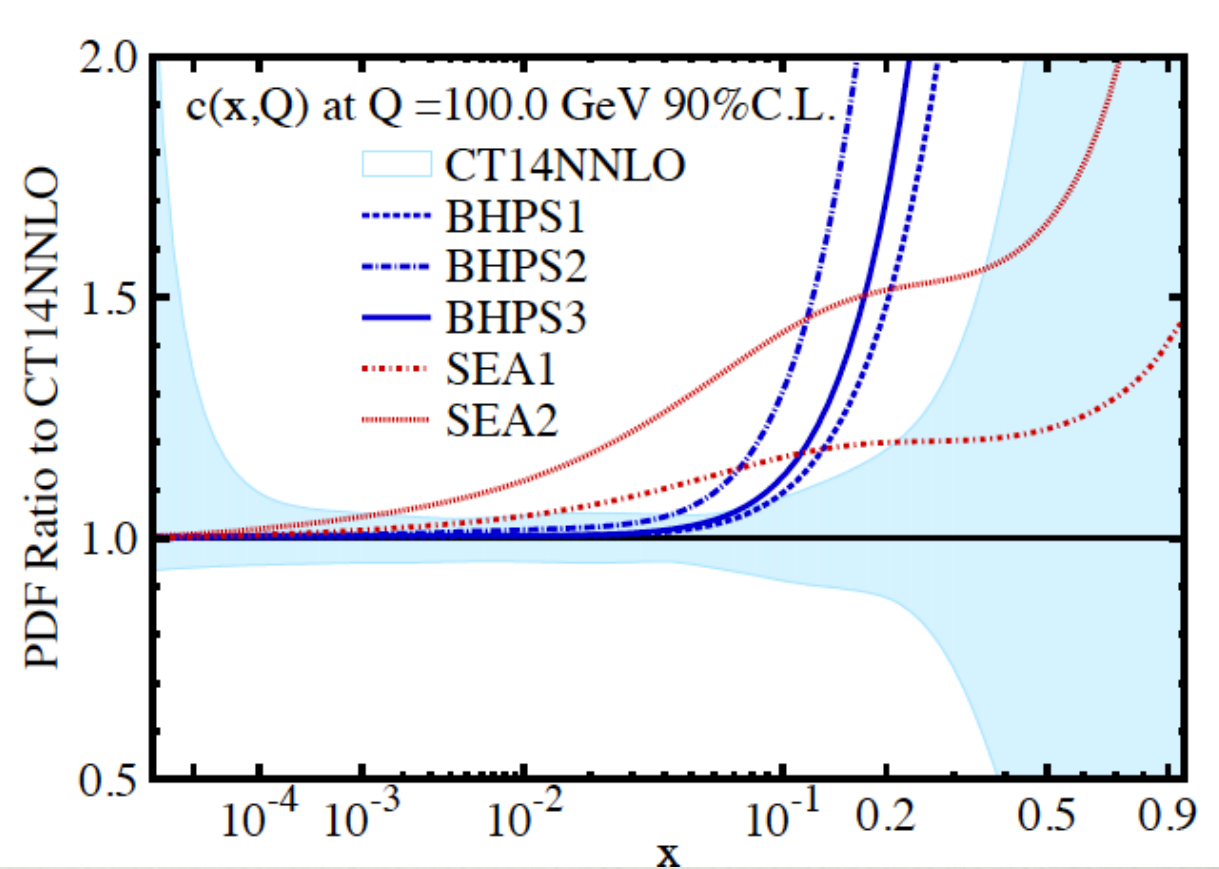}
        {height=0.35\columnwidth}{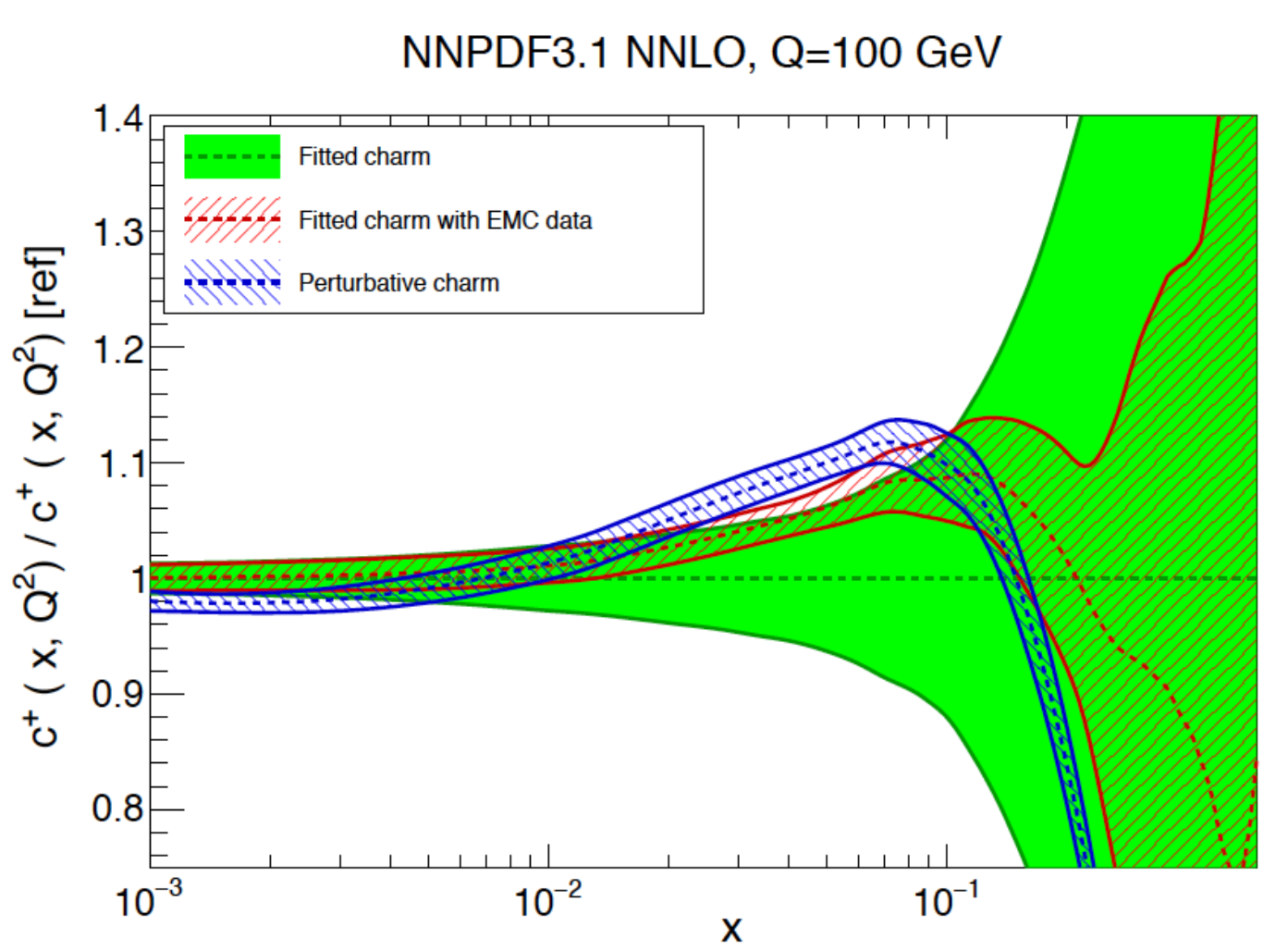}
  \caption{(Left) Ratio of the CT14 charm quark PDF with different
    model fits of intrinsic charm, to the baseline extrinsic only
    PDF~\cite{Guzzi}. (Right) Ratio of the NNPDF fitted charm quark
    PDF, including and excluding the EMC data in the fit, to the
    baseline extrinsic only PDF~\cite{NNPDF}. In both cases the scale
    $Q=100\gev$ is taken.}\label{fig:ic}
\end{figure}

The determination of the intrinsic charm content of the proton was
discussed in~\cite{Guzzi,NNPDF}. In most PDF analyses the charm PDF is
generated extrinsically from perturbative $g\to c\overline{c}$
splittings, and is therefore determined from the other PDFs with no
additional freedom. However, the possibility of some intrinsic
non-perturbative charm content in addition to this extrinsic content
has been a subject of debate for some time. Two different approaches
to this problem were presented here. In a CT
study~\cite{Guzzi,Hou:2017khm} physically motivated models for
\textit{sea}- and \textit{valence}-like components were
considered. While these models determined the intrinsic charm $x$
dependence, the normalisation was left free and fitted to
data. Alternatively, in the NNPDF study~\cite{NNPDF,Ball:2016neh} a
completely agnostic approach was taken and the charm PDF was fitted in
the same fashion as the light quarks. Excluding the early EMC data,
for which questions have been raised with respect to the reliability
of the experimental systematic uncertainties, both studies place a
limit on the momentum fraction carried by intrinsic non-perturbative
charm of $1(2)\%$ for NNPDF (CT). Including the EMC data, NNPDF finds
evidence for non-perturbative charm at the $1.5\sigma$ level. Future
$Z+c$, open-charm, and $J/\psi$ measurements from the LHC should help
further constrain or discover intrinsic charm. The extracted charm
quark PDFs are shown in Fig.~\ref{fig:ic}.

In~\cite{Bonvini} the resummation of $\log^k(x)$ terms in the QCD
splitting functions, which are particularly important at low $x$, was
presented. This was applied for the first time within a PDF fit to a
range of fixed-target and HERA DIS data. Interestingly, some
improvement in the description, in particular at NNLO, was found. The
impact on the PDFs at low $x$ is found to be quite large, and a $\sim
1\%$ change in the Higgs gluon-fusion production cross section is
observed. However, this can only be taken as an indication of possible
effects, as low $x$ resummation should also be included in the
cross-section calculations themselves, which may reduce the
effect. The outlook for including further processes in the context of
a more global fit was discussed, and has subsequently been presented
in~\cite{Ball:2017otu}.

\section{Tools for PDF fits}

In recent years there has been a large increase in theory calculations
for standard-candle processes at NNLO in perturbative QCD. These now
cover, in addition to inclusive $W/Z/\gamma^*$ production, various $2
\to 2$ processes, \textit{e.g.}
$W/Z+\mathrm{jet}$~\cite{Boughezal:2015ded, Ridder:2015dxa},
$t\bar{t}$~\cite{Czakon:2015owf}, and inclusive
jets~\cite{Currie:2016bfm}, in both $ep$ and $pp$ collisions. However,
each calculation requires a large computational effort and thus they
are not directly usable in PDF fits. The \texttt{APPLfast}
project~\cite{Gwenlan} works on extending the grid-technology for NLO
calculations known from \texttt{FastNLO} and
\texttt{APPLgrid}~\cite{Kluge:2006xs,Carli:2010rw} to the calculations
provided by the \texttt{NNLOJET} program. While significant
computations of about $200,000$ CPU-hours must be performed initially,
any subsequent convolution with an arbitrary PDF is completed in much
less than one second. Recently, work on \texttt{FastNLO} tables for
NNLO $t\bar{t}$ calculations has been completed~\cite{Czakon:2017dip}.
The availability of fast NNLO calculations will be an important
ingredient to improve the precision of PDF fits with hadron-collider
data.

The \texttt{APFEL} package~\cite{Bertone:2017gds,Bertone:2013vaa} is a
library for the computation of collinear PDF evolution and DIS
structure functions. In the current version, the package supports
evolution at up to NNLO in QCD and NLO in QED, various options for
heavy-quark contributions (\textit{e.g.} fixed/variable-flavour-number
schemes, pole masses, $\mathrm{\overline{MS}}$ masses), and a fast
computation of DIS observables. \texttt{APFEL} is the main evolution
code used by the NNPDF collaboration~\cite{Ball:2017nwa} and is also
available in the \texttt{xFitter} package~\cite{Alekhin:2014irh,
  Bertone:2017tig}. \texttt{APFEL} features flexible interfaces to
Fortran, C/C++, and Python code.

\begin{figure}[tb]
  \plots{height=0.4\columnwidth}{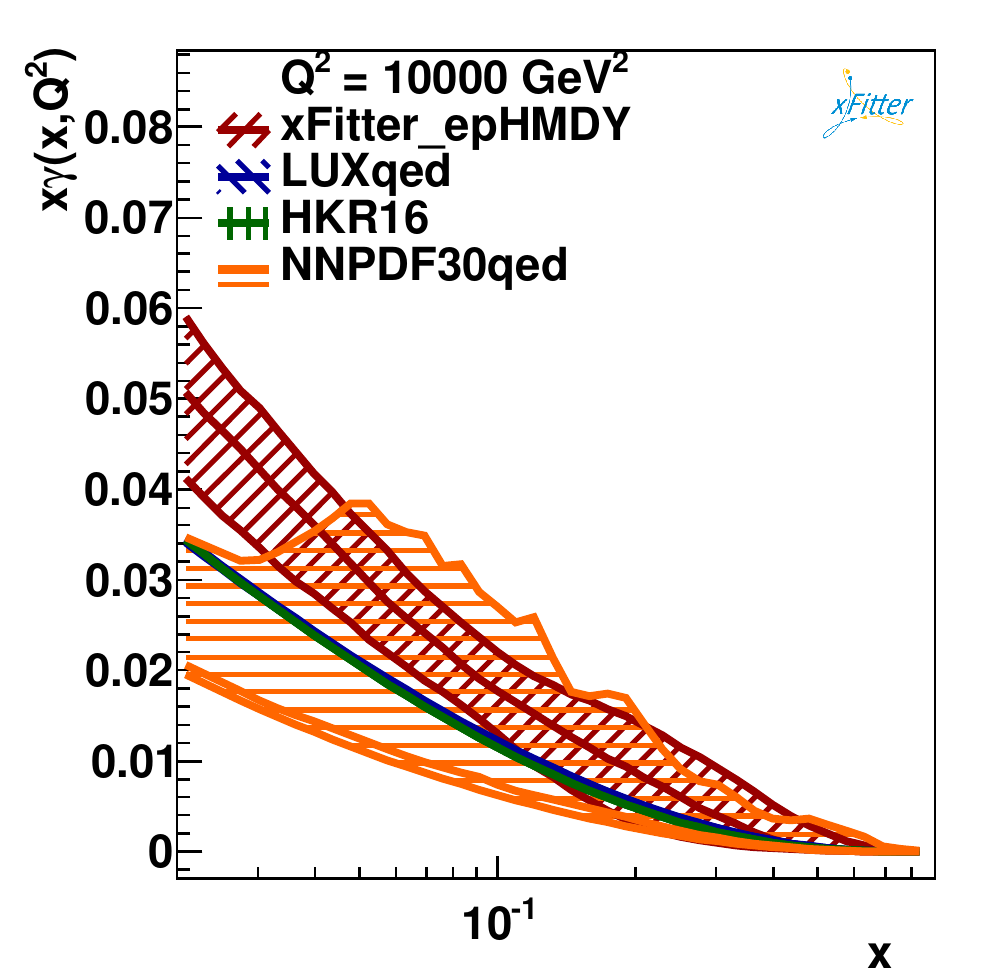}
        {height=0.4\columnwidth}{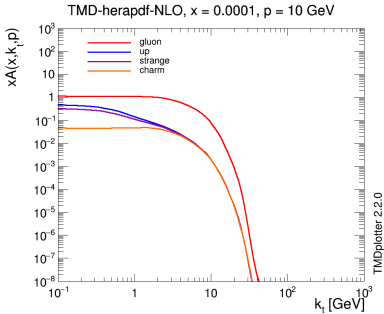}
  \caption{(Left) Extraction of the photon PDF from ATLAS high-mass
    Drell-Yan data~\cite{Giuli:2017tst, Giuli:2017oii}. (Right)
    Transverse momentum dependent PDFs evolved using the parton
    branching model~\cite{Lelek}.\label{fig:xfitterresults}}
\end{figure}

Finally, the \texttt{xFitter} package~\cite{Alekhin:2014irh,
  Bertone:2017tig} is the major platform used for PDF fits and
analyses outside the dedicated PDF collaborations. It is open source
and integrates several other programs that are necessary to perform a
PDF analysis. The \texttt{xFitter} team has recently developed a new
major release, 2.0.0, with significant improvements. There are more
than 30 public results obtained with \texttt{xFitter} since the start
of the project. Members of the LHC experiments are currently the main
developers and users of \texttt{xFitter}, using the program to assess
data sensitivity to PDFs and compatibility with QCD predictions. Two
recent results by the \texttt{xFitter} team were featured at this
workshop. The first was an extraction of the photon PDF using
high-mass Drell-Yan data from the ATLAS experiment at the LHC
~\cite{Giuli:2017tst, Giuli:2017oii}. The second result explored a new
approach to performing the evolution of transverse momentum dependent
PDFs using the parton branching model~\cite{Lelek}. These two results
are showcased in Fig.~\ref{fig:xfitterresults} (left) and (right),
respectively.

\section{PDFs and effects in heavy Nuclei}

The study of PDFs for heavy atomic nuclei continues to be an
interesting topic with a steady level of activity. As has been known
for some time, a nucleus is more than the sum of free nucleons, due to
nuclear dynamics. Recent new data on heavy ion collisions have
increased both the interest in predictions and expanded the available
datasets. Traditionally $\nu N$ scattering has been an important
source of information on the strange-quark density, which requires an
understanding of the nuclear corrections, as discussed
in~\cite{Olness}.  While the EMC effect has been known since 1983,
work is ongoing to describe this theoretically as a combination of
several effects, including dynamic short range correlations, as
discussed in~\cite{Hen}.

\begin{figure}[tb]
  \begin{center}
    \includegraphics[width=.95\linewidth]{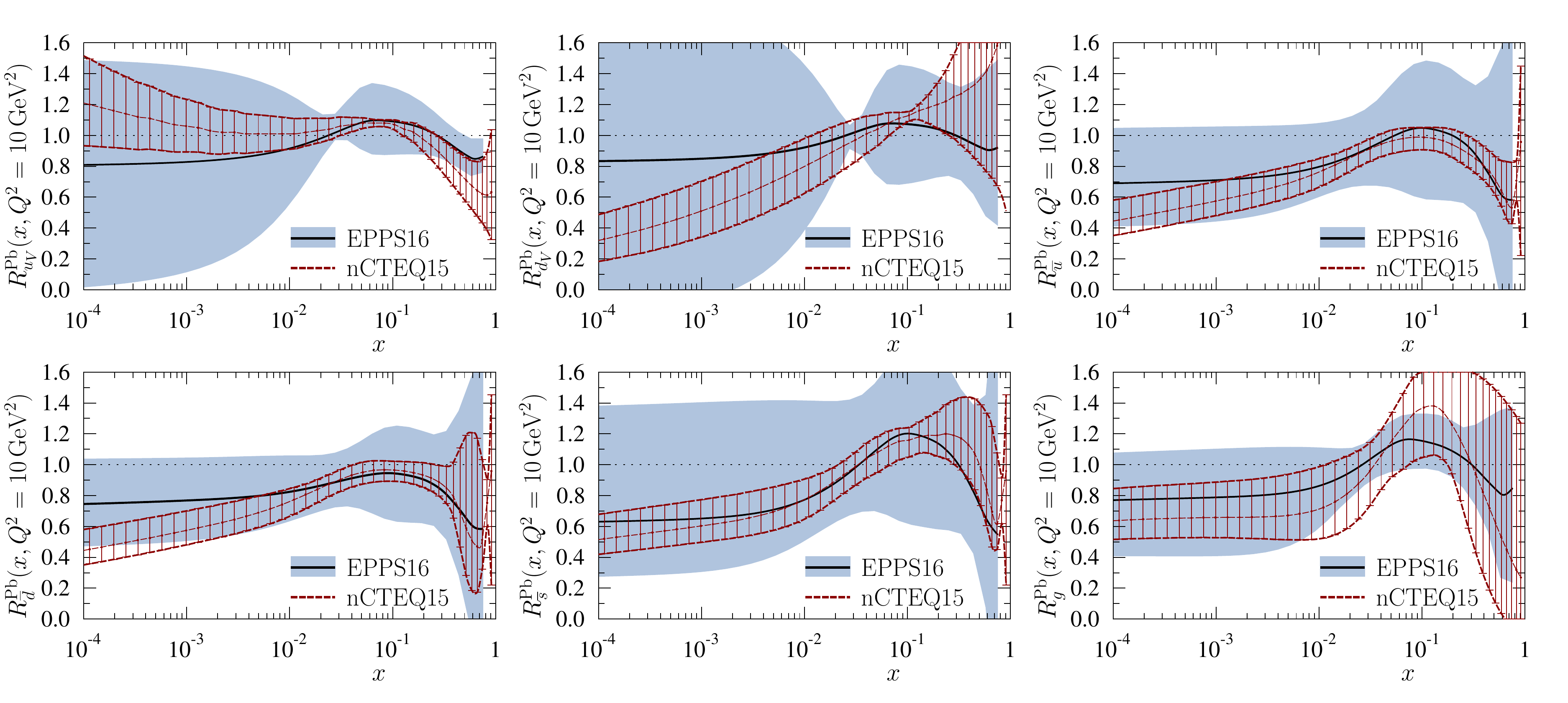}
  \end{center}
  \caption{A comparison of the nuclear correction functions for
    $\mathrm{Pb}$ nuclei as determined in the recent nCTEQ15 and
    EPPS16 analyses~\cite{Eskola:2016oht}.\label{fig:epps16}}
\end{figure}

The extraction of nuclear PDFs typically proceeds with a model to
parameterise the nuclear effects. The new EPPS16 fit was presented
in~\cite{Eskola:2017rmp, Eskola:2016oht, Arleo:2017jgn}, which relies
on a parametrisation of these effects that is fitted to data. For the
first time, this fit includes LHC data, which has allowed the
parametrisation to be more flexible to reduce the impact of initial
assumptions.  The inclusion of pion-nucleus Drell-Yan data show some
sensitivity to nuclear valence-quark modifications. The approach by
the nCTEQ collaboration was presented in~\cite{Olness}, which features
a generalised $A$-dependent parameterisation of nuclear
effects. Fig.~\ref{fig:epps16} shows an overlay of the most recent
EPPS and nCTEQ results regarding the modification factor for lead
nuclei.  A general agreement on the central values can be
seen. However, the uncertainties are large and their estimation is
typically driven by model choices rather than the data.  The
Kulagin-Petti approach was presented as well in~\cite{Petti}, which
uses a microscopic model incorporating Fermi motion.

\begin{figure}[tb]
  \plots{height=0.45\columnwidth}{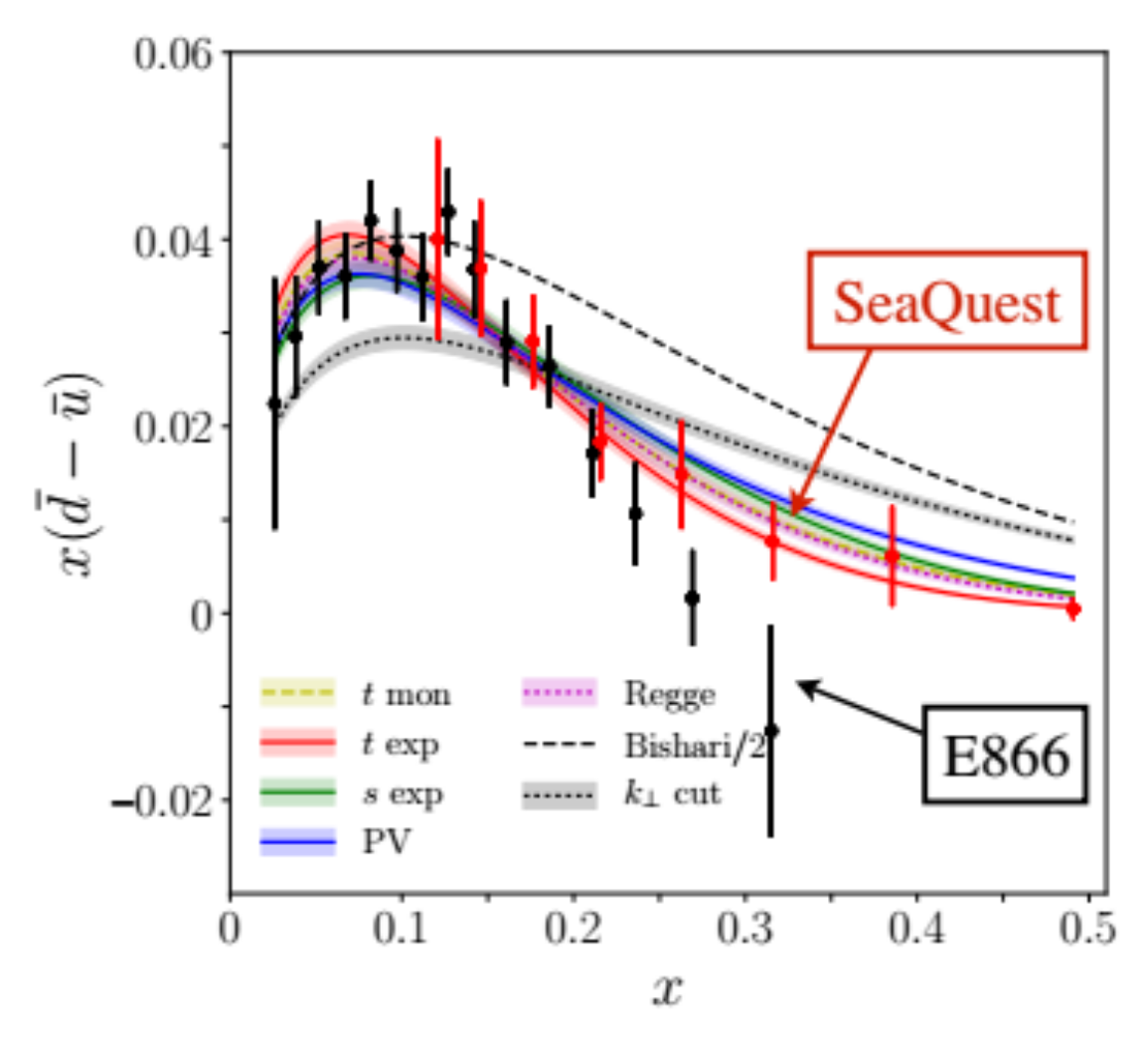}
        {height=0.45\columnwidth}{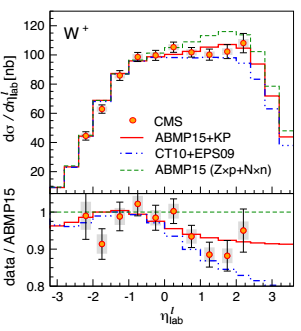}
  \caption{(Left) Data from the E866 and SeaQuest experiments on
    $\bar{d}-\bar{u}$~\cite{Melnitchouk}. (Right) An example of
    $p\mathrm{Pb}$ LHC data, here $W^+$ production, that is used to
    differentiate between nuclear models and constrain model
    parameters~\cite{Petti}.\label{fig:nucleardata}}
\end{figure}

Another topic of continued interest is the combination $\bar{d} -
\bar{u}$ at $x \approx 0.1$, which was found to be significantly
larger than zero by the E866 Drell-Yan experiment. Explanations for
this effect were discussed in~\cite{Melnitchouk,
  Paakkinen:2017eat}. These typically use models based on pion
exchange, which can be obtained from QCD and calculated in chiral
effective field theory. New data by the SeaQuest collaboration has
recently disfavoured a possible change of sign change at $x > 0.3$, as
shown in Fig.~\ref{fig:nucleardata} (left). In the future, leading
neutron analyses and Drell-Yan data with pion beams will constrain
pion PDFs at low and high $x$ in an upcoming neutron tagged DIS
experiment at JLAB~\cite{Keppel}.

Several LHC measurements sensitive to nuclear effects were presented
by the experiments~\cite{Mischke, Chapon} or discussed further by the
fit collaborations~\cite{Eskola:2017rmp, Petti, Olness}. These
included most prominently data on the production of jets or $W/Z$
bosons in LHC $p\mathrm{Pb}$ collisions. Comparisons of jet production
between $pp$ and $p\mathrm{Pb}$ collisions have been made, and while
they are similar when integrated in rapidity, CMS data show
significant modification as a function of jet rapidity. Indeed, this
data already provides some constraints on the EPPS16 fit. An example
of data on $W^+$ production in $p\mathrm{Pb}$ collisions compared to a
variety of predictions is shown in Fig.~\ref{fig:nucleardata} (right);
not all nuclear modifications improve the description of the $W^\pm$
and $Z$ data. Similar to the high precision $pp$ case, correlations in
these $p\mathrm{Pb}$ data are starting to be exploited, \textit{e.g.}
by analysing the forward/backward and $W^+/W^-$ asymmetries. More
precise measurements in the near future will significantly improve our
knowledge. New $\nu N$ cross-section data is expected in the near
future from the MINERvA experiment~\cite{Wospakrik}, although the reach
in $Q^2$ will be limited.

\section{New Theory Developments}

\begin{figure}
  \plots{height=0.4\columnwidth}{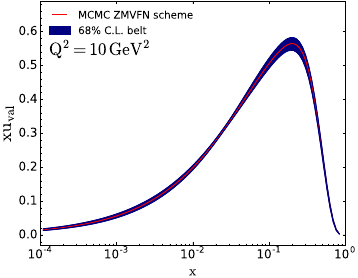}
        {height=0.4\columnwidth}{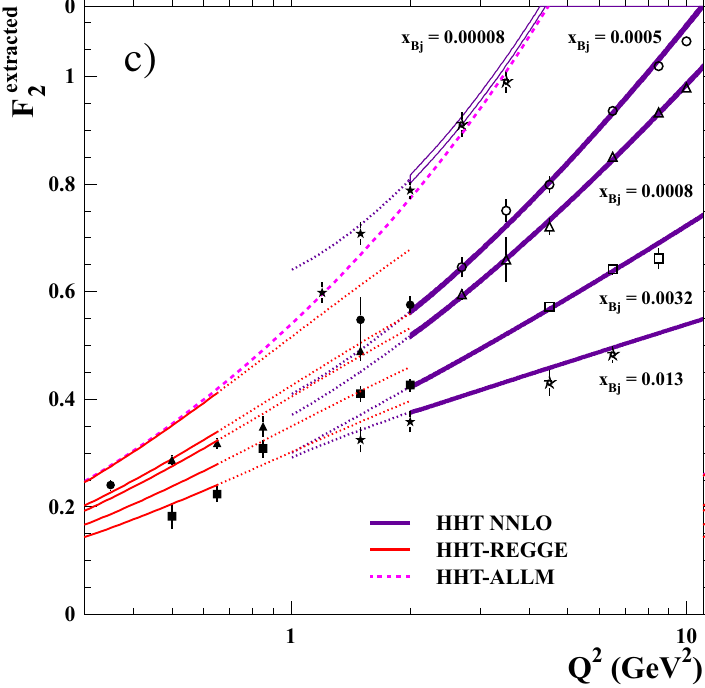}
  \caption{(Left) Up valence quark distribution extracted using the
    Markov Chain Monte Carlo technique~\cite{Mariane}. (Right) The
    structure function $F_2(Q^2)$ for selected values of $x_{Bj}$,
    compared to the Regge and NNLO DGLAP based
    predictions~\cite{Abt:2017nkc}.\label{fig:mcmcnat}}
\end{figure}

A number of new theoretical ideas and developments were also discussed
at the conference. In~\cite{Nocera:2017war} a summary of the first
joint PDF and lattice QCD workshop, which took place in Oxford, March
2017, was presented. The aim is to explore the possibility for lattice
QCD calculations to be used as constraints for global PDF fits. Many
topics were discussed at this workshop, including the need to identify
benchmarks that are well constrained by global fits and can therefore
be used to validate the lattice inputs, the choice of observables that
could benefit from lattice constraints, and the understanding of
various sources of systematic uncertainty in the lattice calculations.
This has subsequently been summarised in the form of a white
paper~\cite{Lin:2017snn}, which serves as a basis for future work in
the area.

In~\cite{Mariane,Gbedo:2017eyp}, a novel approach to performing PDF
fits was described. Markov Chain Monte Carlo techniques were used to
extract the probability density functions of PDFs, allowing for
further information on the PDF uncertainties. The minimisation was
performed using a hybrid (or Hamiltonian) Monte Carlo technique, first
applied in lattice QCD applications. The implementation used the
\texttt{xFitter} package and a validation was performed using the HERA
DIS dataset with the zero-mass variable-flavour number scheme. The
results were found to be consistent with the corresponding HERAPDF
set, thus serving as a proof-of-concept. The extracted $u$-valence is
shown in Fig.~\ref{fig:mcmcnat} (left).

In~\cite{JunMC,Hou:2016sho} the determination of CTEQ-TEA MC replica
sets was discussed. Alternative techniques for converting from the
Hessian to MC PDFs were considered, either sampling directly in the
PDF $f$, or in $\ln f$. The latter case guarantees PDF positivity,
provided the original Hessian set is positive within its
uncertainty. In addition, a shift is introduced to guarantee that the
central value of the MC replicas corresponds exactly to the central
Hessian set, something which is otherwise not guaranteed. The MC sets
using these techniques were compared to the original Hessian PDFs, and
found to agree well. These are now publicly available in LHAPDF
format.

In~\cite{Davies:2017azj,Davies:2016jie}, a new calculation of certain
contributions to the 4-loop splitting functions was discussed. In
particular, a lattice basis reduction technique was applied to
reconstruct the large $n_F$ contributions to the splitting functions
from the calculated Mellin moments. These results have been validated
against a range of known results in various limiting regions, and are
an important step towards a consistent treatment of PDFs at N$^3$LO.

The description of HERA DIS data in the low $x$ and $Q^2$ region was
discussed in two
talks~\cite{Motyka,Wichmann}. In~\cite{Wichmann,Abt:2017nkc} the HERA
data was fit using a two-component approach, with standard
perturbative QCD DGLAP applied above a $Q^2$ of $2\gev^2$, and a
Regge-inspired model applied below. The data agrees well with these
approaches in their expected regions of validity, but not in the
transition region. However, as shown in Fig.~\ref{fig:mcmcnat} (right)
the data shows a smooth transition between the two regions, prompting
the conclusion that nature does not ``know'' about perturbation
theory. In~\cite{Motyka,Motyka:2017xgk} a QCD-inspired model of
higher-twist corrections was used to supplement standard DGLAP
evolution, with the parameters of these corrections determined in a
simultaneous fit to the HERA data. A large improvement in the fit
quality at low $Q^2_{\rm min}$ was found upon the inclusion of these
corrections, providing possible evidence for the important
contribution of higher twists at HERA.

\section{Conclusions}

The structure of hadrons, specifically PDF dynamics, is an interesting
topic that continues to attract attention from the larger nuclear and
particle physics communities. From a practical perspective,
experiments at the current and future energy frontiers using hadron
beams need PDF information to exploit their data. Without improved DIS
data, much of the attention has now turned to data from the LHC
experiments, with a continuous effort to measure and predict
standard-candle processes at the highest possible precision, and
include these data into global PDF fits. Many of the talks in this
working group revolved around these topics.  For the case of heavy
nuclei the range of datasets is more scarce than for protons, giving
the new data from LHC and elsewhere a more prominent role.

Beyond the topic of the LHC data and applications, fundamental topics
raise interest, including the validity of the framework of collinear
factorisation, the treatment of heavy-quark contributions,
non-perturbative effects and techniques, resummation effects, and the
nature of PDFs at very small or large $x$.

Many of these topics were debated lively in the working groups
sessions, demonstrating that the DIS conference is one of the focal
points of the community for these discussions.

\section*{Acknowledgements}

We would like to thank all the speakers and participants that
contributed to the success of the working group 1 programme with
interesting presentations and lively discussions. We also thank the
organisers for a well-organised and interesting DIS 2017 conference,
and for giving us the opportunity to convene this session.

\bibliographystyle{h-physrev}
\bibliography{main}{}

\end{document}